\documentclass[10pt,conference,onecolumn]{IEEEtran}
\usepackage{amsmath,amsfonts}
\usepackage{algorithmic}
\usepackage{array}
\usepackage[caption=false,font=normalsize,labelfont=sf,textfont=sf]{subfig}
\usepackage{textcomp}
\usepackage{stfloats}
\usepackage{url}
\usepackage{verbatim}
\usepackage{graphicx}
\def\BibTeX{{\rm B\kern-.05em{\sc i\kern-.025em b}\kern-.08em
    T\kern-.1667em\lower.7ex\hbox{E}\kern-.125emX}}
\usepackage{balance}
\usepackage{tabularx}
\usepackage{lmodern}
\usepackage{comment}
\usepackage{booktabs}
\usepackage{multirow}
\usepackage[greek, english]{babel}
\DeclareUnicodeCharacter{03BC}{\ensuremath{\mu}} 
\usepackage{fancyhdr}

\begin{document}

\bstctlcite{IEEEexample:BSTcontrol}

\title{TriCloudEdge: A multi-layer Cloud Continuum}

\author{George Violettas, %
\thanks{G. Violettas is with SYSGO GmbH, Germany, and a member of the SWNRG research lab at the University of Macedonia, Greece.}%
\and Lefteris Mamatas ($\dagger$) \thanks{L. Mamatas was a professor and head of the SWNRG research lab at the University of Macedonia, Greece. \textbf{Sadly, Professor Mamatas is deceased (October 2025).}}}

\markboth{TriCloudEdge: A multi-layer Cloud Continuum}{TriCloudEdge: A multi-layer Cloud Continuum}

\maketitle

\footnotetext[1]{G. Violettas is with SYSGO GmbH, Germany, and a member of the SWNRG research lab at the University of Macedonia, Greece.}
\footnotetext[2]{L. Mamatas was a professor and head of the SWNRG research lab at the University of Macedonia, Greece. \textbf{Sadly, Professor Mamatas is deceased (October 2025).}}

\pagestyle{fancy}
\fancyhf{}
\fancyhead[C]{\normalsize TriCloudEdge: A multi-layer Cloud Continuum}
\fancyfoot[C]{\thepage}
\fancyfoot[L]{\footnotesize Preprint --- submitted to arXiv}
\renewcommand{\headrulewidth}{0.4pt}
\renewcommand{\footrulewidth}{0.4pt}
\fancypagestyle{plain}{%
  \fancyhf{}
  \fancyfoot[C]{\thepage}
  \fancyfoot[L]{\footnotesize Preprint --- submitted to arXiv}
  \renewcommand{\headrulewidth}{0pt}
  \renewcommand{\footrulewidth}{0.4pt}
}

\begin{abstract}

TriCloudEdge is a scalable three-tier cloud continuum that integrates far-edge devices, intermediate edge nodes, and central cloud services, working in parallel as a unified solution. At the far edge, ultra-low-cost microcontrollers can handle lightweight AI tasks, while intermediate edge devices provide local intelligence, and the cloud tier offers large-scale analytics, federated learning, model adaptation, and global identity management. 
The proposed architecture enables multi-protocols and technologies (WebSocket, MQTT, HTTP) compared to a versatile protocol (Zenoh) to transfer diverse bidirectional data across the tiers, offering a balance between computational challenges and latency requirements. 
Comparative implementations between these two architectures demonstrate the trade-offs between resource utilization and communication efficiency. The results show that TriCloudEdge can distribute computational challenges to address latency and privacy concerns. 
The work also presents tests of AI model adaptation on the far edge and the computational effort challenges under the prism of parallelism. This work offers a perspective on the practical continuum challenges of implementation aligned with recent research advances addressing challenges across the different cloud levels.

\end{abstract}

\begin{IEEEkeywords}
Cloud, IoT, Cloud Continuum, Edge Cloud, Edge AI, WebSocket, MQTT, Zenoh
\end{IEEEkeywords}

\section{Introduction}



\IEEEPARstart{T}{he} rapid evolution of the Internet of Things (IoT) in recent years and the consequent increase in demand for real-time intelligent decision-making close to the data generation sources have brought a powerful change in the Artificial Intelligence (AI) deployment, i.e., moved it closer to the primary event areas and space. 
The above challenges led to the emergence of the Edge-Cloud AI Continuum, a disturbing distributed architecture that integrates, in harmony, decentralized constrained edge resources with central cloud infrastructure to offer scalable, real-time, low-latency, privacy-preserving, end-to-end, AI-capable services.
The continuum is not only an improvement of decentralized services and close-to-the-source applications but also a new dynamic in workload placement and an adjusted and adapted way of fully utilizing the potential and specific characteristics of each device while allocating AI services on each level as needed.

Orchestrating this so-called Edge-Cloud AI continuum reveals the necessity of integrating multiple diverse communication technologies, like ''WebSocket'' for handling edge-to-far-edge links and custom protocols for big data fragmentation and standard cloud protocols like Message Queuing Telemetry Transport (MQTT), not to exclude the very basic HTTP for communication from the edge to the cloud and back~\cite{IoTSurvey2025}. Such a multi-protocol approach inevitably introduces overhead on resource consumption, especially on constrained devices, e.g., far edge and edge~\cite{DJAMAA2025104357}, along with the maintenance and even development overhead (compile time is a serious drawback here), affecting the dataflow, the optimization, and the effectiveness across the whole continuum.

The work at hand focuses on the capabilities and exploitation of Edge-AI within the above-described cloud continuum, as well as the execution of AI models at the edge and the subsequent issues and challenges that such an endeavor can reveal. Recent advantages of Edge-AI, such as ultra-low-latency, reduced bandwidth demand, and enhanced security and autonomy, can now be utilized by locally operating applications to make decisions and act upon them. 
A critical advantage of the continuum is the locality of the information. By processing sensitive data, such as human faces or event triggers, directly at the far-edge or edge tiers, the system minimizes the exposure of such data. Hence, locality increases the resilience against network outages and reduces the attack surface, since it is possible to further transmit aggregated or anonymized results to higher or external layers. The system aligns with privacy-preserving and regulatory compliance. 
The practical implementations of the continuum are vast, based on the economies of scale and the capability of distributing AI tasks over diverse hardware devices, adjusted to different needs and use cases. 

For example, at the Far Edge, highly constrained devices can perform initial, lightweight AI tasks such as face detection. At the next level, i.e., the Edge, more (AI-) capable devices can handle more complex local inference, such as identifying an already detected face against a local database. Finally, the Cloud layer, enabled by globally provided cloud services, provides abundant computational power and storage capacity for advanced analytics, large-scale model training, and federated learning orchestration when needed. 
This three-tier approach offers resource allocation optimization and utilization, making sure that computational costs are distributed efficiently and, hence, subsequently maximizing performance while keeping the operational cost low.

Although we can find today several works describing edge, fog, and cloud architectures, there is still a lack of end-to-end implementation and evaluation to demonstrate how AI workloads can be distributed among the far-edge, edge, and cloud tiers under real hardware constraints. Existing approaches are either conceptual, simulation-only, or focused on one of those tiers. Hence, the holistic approach of the evaluation of a three-tier continuum over heterogeneous constrained devices, utilizing diverse communication protocols and realistic AI tasks, remains open.

This paper provides a comprehensive analysis of the architectural principles, capabilities, and data flow within the Edge-Cloud AI Continuum, along with the presentation of two applicable, fully functional reference implementations of the three-layer Cloud Continuum and a detailed comparison between them, as presented in Section~\ref{sec:system_architecture}. To that end, we implemented two comparative architectures, i.e., a Multi-Protocol Architecture (based on WebSocket) and a Zenoh-Unified Architecture, and compared them under identical workloads, as described in Section~\ref{sec:comp_analysis}.

The main contributions of this work are: (1) the design and implementation of the TriCloudEdge three-tier Continuum on real hardware; (2) the implementation and comparison of two communication architectures, i.e., multi-protocol (WebSocket, HTTP, MQTT) and Zenoh-Unified, under identical AI workloads; and (3) a quantitative evaluation of latency, throughput, and pipeline parallelism using applied figures and metrics, reported in detail in Section IV.

\subsection{Edge-AI Global Initiatives}
Edge-AI has recently become a strategic priority for both European and global ecosystems. Towards that goal, there are targeted calls from Horizon Europe~\cite{Horizon2025EdgeCloud}, initiatives from the World Economic Forum~\cite{WEF2025EdgeAI}, and a massive investment through public-private partnerships and industrial alliances (e.g., Edge-AI Foundation~\cite{EdgeAIFoundation2025}), aiming to embed AI capabilities directly into edge devices, enabling responsive applications in sectors like autonomous mobility, smart manufacturing, and precision agriculture.
Moreover, a recent EU study clearly mentions that the proliferation of compute-intensive Generative AI (GenAI) models shows that deploying large-scale AI solely to the cloud is neither economically nor environmentally sustainable, and hence there must be a substantial effort towards reducing latency, ensuring data privacy, and optimizing consumption of energy by pushing processing capabilities to the edge~\cite{EPoSSINSIDE_EdgeAI_2025}.

Such deployments of Edge-AI have to rely upon the efficient and distributed resource-orchestration across a multi-layer cloud continuum, maximizing the utilization of each level's special characteristics and capabilities. Such an architecture would execute lightweight inference tasks on far-edge constrained devices (e.g., wearables). More advanced and complex analytics could take place at the next level, i.e., the edge. Lastly, computationally intensive multiple models and even federated learning or global search could take place in the central cloud.

The European Union recognizes the strategic significance of Edge-AI by funding such end-to-end multi-layer approaches for edge-cloud deployments, underlining the need for scalable and low-latency AI services~\cite{Horizon2025EdgeCloud}. Other recent global initiatives, such as the Edge-AI Continuum~\cite{LatentAI2025EdgeContinuum}, dictate that only a distributed, hybrid architecture can satisfy the latency, privacy, and energy efficiency needs of the current AI applications. Furthermore, without the continuum topology, Edge-AI might become rather unsustainable and even fragmented, since isolated edge nodes do not have the capacity to do model training, secure data aggregation, or exercise federated learning. Hence, the continuum is not only a technical achievement but also a strong prerequisite for supporting Edge-AI~\cite{mishra2024systematic} to the maximum potential of it.

\begin{table}[htbp]
\caption{Comparative Analysis of Edge-AI Challenges from Selected Papers}
\label{tab:challenges}
\centering
\renewcommand{\arraystretch}{1.2}
\small
\begin{tabular}{p{2.5cm}p{2.6cm}p{2.6cm}p{2.6cm}p{2.6cm}}
\hline\hline
\textbf{Major Challenge Category} & \textbf{Tobias Meuser et al~\cite{revisiting2024}} & \textbf{Vijay Janapa Reddi~\cite{acm25generative}} & \textbf{Gill et al.~\cite{gill2025taxonomy}} & \textbf{Vasuki Shankar~\cite{shankar2024edge}} \\
\hline
\textbf{Resource Constraints} & Model architecture, power budgets & Compute, memory, energy, thermal & Computational capabilities & Computational power, memory, energy \\
\hline
\textbf{Data Management} & Data privacy, network load for training data & Scarcity, quality, fine-tuning, privacy, governance & Privacy & Privacy, security \\
\hline
\textbf{Model Optimization \& Reliability} & Model architecture constraints & Compression, hallucinations, alignment, safety, continual learning & Model optimization techniques & Model size, inference speed, power consumption \\
\hline
\textbf{Security \& Trustworthiness} & Secure distribution/execution, privacy by architecture & Privacy preservation, data governance & Vulnerabilities, privacy & Threats, adversarial attacks, explainability, virtualization layer \\
\hline
\textbf{Network \& Interoperability} & Volatile transport links, network load & Federated learning overhead & Scalability & Interoperability, compatibility, frequent communication \\
\hline
\end{tabular}
\end{table}

\section{Related Work}
\label{sec:related_work}

\begin{figure}[!t]
\centering
\includegraphics[width=\textwidth]{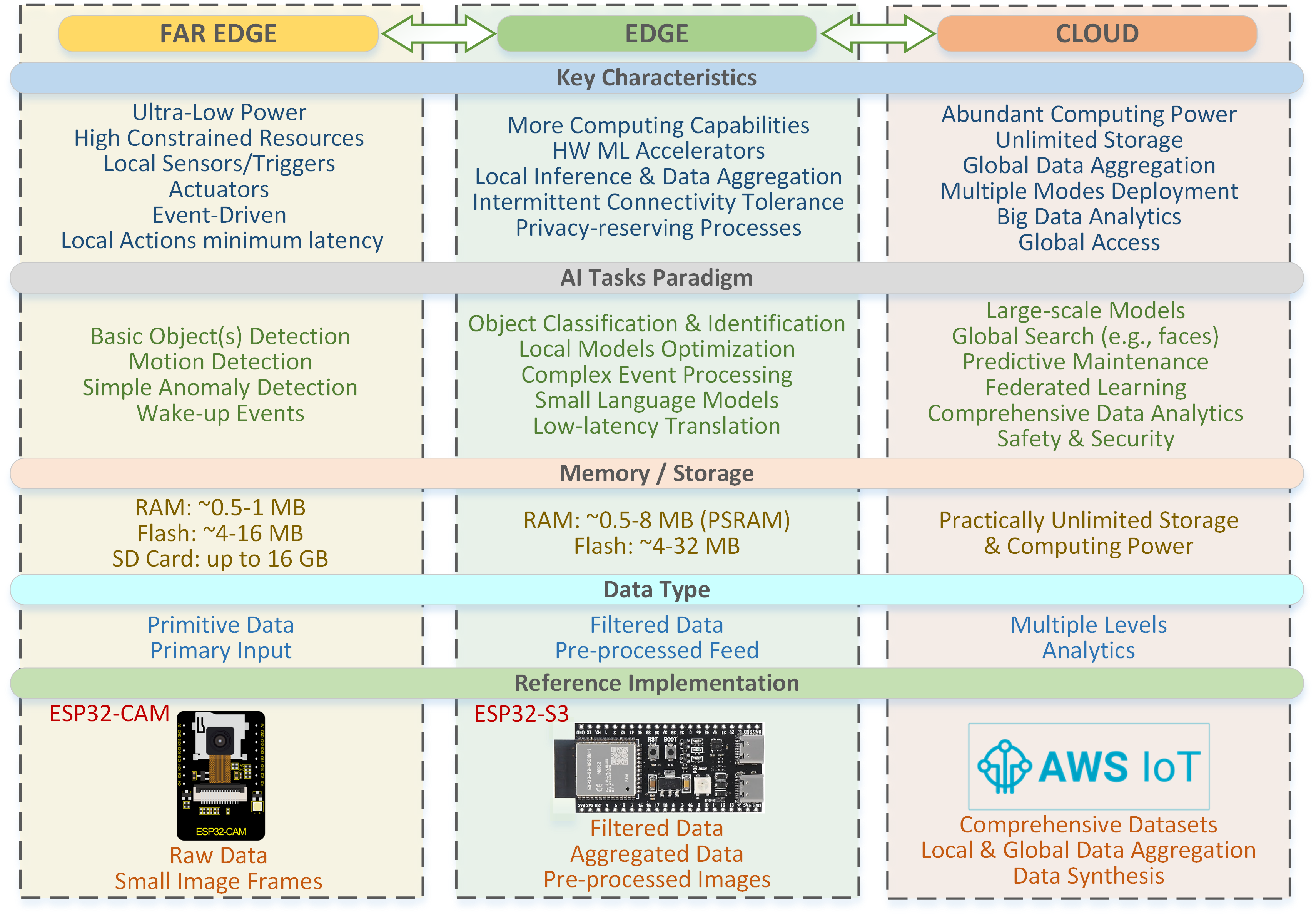}
\caption{Edge-Cloud AI Continuum}
\label{fig:Continuum}
\end{figure}

\subsection{Edge Cloud Terms and Status}

The idea of edge computing starts with the notion of ``fog computing'', first quoted by CISCO~\cite{cisco2012}, as the extension of cloud computing to the edge of the network as an enabler of new applications and services.

In~\cite{cloudlets2009}, the term ``cloudlets'' was introduced as the elements in-between resource-constrained mobile devices and the distant but resource-rich cloud. The authors state that although the mobile hardware keeps on evolving (following Moore's law), it will always be resource-poor compared to more centralized and static ones. With the exception of mobile phones, this is still true today, as is, for example, the case for wearable devices. Also in~\cite{edge2016vision}, the term ``Edge'' is described as allowing computation at the edge of the network, on downstream data from cloud services and upstream data from the IoT (Internet of Things) services. Furthermore, the authors describe a gateway as the edge between home things and the cloud, i.e., three levels of cloud(s), in a direct analogy with the topic of the paper at hand.

The ETSI (European Telecommunications Standards Institute) organization specified Mobile Edge Computing (MEC) as an ICT technology with cloud-computing capabilities at the edge of the mobile network and near mobile subscribers, targeting to reduce the latency and ensure highly efficient network operation and services delivery~\cite{mec2015}. Furthermore, MEC pushes network control and storage functions to the edge, enabling constrained devices to perform computation-intensive and latency-critical tasks~\cite{mao2017survey}. 

The above notions led to the evolution of the Continuum paradigm, sometimes named "computational hierarchy"~\cite{2022continuum}, integrating three distinct layers, i.e., Far Edge (highly constrained devices), Edge (intermediate nodes with advanced capabilities), and the Cloud (resource-abundant data centers). This multi-tier architecture is capable of supporting modern, decentralized AI applications by balancing processing power and responsiveness across the entire network~\cite{caiazza2022edge}.

\subsection{Advantages of Edge-AI}

Regarding implementing AI at the edge, some recent works~\cite{shankar2024edge} also describe it as Distributed AI (DAI). End-edge-cloud computing (EECC)~\cite{duan2022distributed} mentions that it has several advantages, such as bandwidth efficiency, resilience, and reduced latency~\cite{shankar2024edge, duan2022distributed}, improved security and privacy, enhanced scalability, resilience, and reliability, along with greater context awareness~\cite{shankar2024edge}. Applications span from smart homes to industrial IoT~\cite{duan2022distributed}, including smart-city energy systems that use IoT intelligence for optimization~\cite{nikpour2025intelligent}.

Other attractive factors include real-time response, delivering near-instantaneous feedback to augmented reality applications, robotic systems, smartwatches predicting stress levels, and medical devices~\cite{acm25generative}.

\subsection{Edge-AI Issues and Challenges}
Some significant challenges faced by Edge-AI include resource constraints, e.g., limited computational power and memory capacity~\cite{DJAMAA2025104357}, and energy efficiency~\cite{shankar2024edge}.

The necessity to create architecture-constrained designs to address the profoundly limited resources of edge devices, such as the secure distribution and execution of AI models in conjunction with the substantial network load required for distributing models, is a real challenge, with or without collecting data for training~\cite{revisiting2024}.
An obvious primary challenge is that state-of-the-art generative models cannot fit or run on such far-edge constrained devices~\cite{revisiting2024}. Also, scalability remains a major challenge~\cite{gill2025taxonomy}.

Trying to optimize the Edge-AI models, extensive considerations and careful planning are needed over factors like model size, inference speed, and not to neglect power consumption~\cite{revisiting2024, shankar2024edge}.

Moreover, a negative factor towards adoption within the security domain specifically is the lack of explainability, along with the complexity of the design and maintenance required by the virtualization layer for distributed deployment~\cite{shankar2024edge}. 
Other factors include the quantization and pruning for data security and privacy, usually revealing existing security vulnerabilities~\cite{gill2025taxonomy}.
The authors at~\cite{acm25generative} refer to the issues that Edge-AI currently faces, grouped as major challenges identified and organized into three interacting dimensions, i.e., data, model, and compute.

A valid taxonomy of Edge-AI could include categories like application architecture, resource management, and methodologies~\cite{gill2025taxonomy}.
A synopsis of the major challenges that Edge-AI is facing, based on selected recent papers, is presented in Table~\ref{tab:challenges}.

Yet, there is still limited work on end-to-end implementation and evaluation of a three-tier continuum on real constrained hardware with a direct comparison of multiple different communication protocol implementations. This paper addresses that gap by presenting TriCloudEdge, a deployed three-tier continuum with two communication architectures (Multi-Protocol and Zenoh-Unified), and by evaluating latency, throughput, and parallelism, as described in Section~\ref{sec:comp_analysis}.

\begin{table}[htbp]
\label{table:cloud-architecture}
\caption{Far Edge, Edge, and Cloud Architectures}
\centering
\renewcommand{\arraystretch}{1.2}
\begin{tabular}{p{2.4cm}p{3.2cm}p{3.2cm}p{3.2cm}}
\hline\hline
\textbf{Aspect} & \textbf{Far Edge (Ultra-Constrained)} & \textbf{Edge (Intermediate)} & \textbf{Cloud (High-Capacity)} \\
\hline
Computational Power & Very low (e.g., microcontrollers, like ESP32) & Moderate (e.g., gateways, edge servers like ESP32-S3) & High (cloud providers like Amazon Web Services, AWS) \\
\hline
Storage Capacity & Kilobytes to Megabytes & Megabytes to Gigabytes & Terabytes to Petabytes \\
\hline
Energy Source & Battery-powered, low energy budget & Limited (e.g., solar, PoE) & Mains/grid powered \\
\hline
Connectivity & Intermittent (LoRa, BLE) & Stable (Wi-Fi, 5G, Ethernet) & High-speed backbone (fiber optic) \\
\hline
Latency Sensitivity & Ultra low (local, near real-time) & Low to moderate & Higher latency well handled \\
\hline
Security & Basic hardware security, physical access & Secure boot, software-level controls & Full-stack, policy-driven security \\
\hline
Data Processing & Minimal/local preprocessing & Real-time analytics & ML training, large-scale analytics \\
\hline
Deployment Location & On-device or near device & Local infrastructure & Remote/cloud data centers \\
\hline
Update Frequency & Rare; Restrictions on over-the-air updates & Regular updates feasible & Frequent CI/CD, DevOps deployments \\
\hline
Example Devices & Sensors, Wearables & Smart Cameras, Autonomous Vehicles & Cloud Storage  Services, AI platforms \\
\hline
\end{tabular}
\end{table}

\section{System Architecture \& Reference Implementation} %
\label{sec:system_architecture}

To address the gap identified in Section~\ref{sec:related_work}, we implemented a three-tier Edge Continuum with two communication architectural designs, described and compared in Section~\ref{sec:comp_analysis}.
The system presented in this work can be described as an ``Edge Continuum'' over three cloud levels or tiers. The basic notion is that the first level, i.e., the ``Far Edge'', acts as the tier very close to the data generator or any triggering event, like basic shape, motion, or even face detection. 

The second level, i.e., the ``Edge'', is where the ``Far Edge'' sends filtered or aggregated data for further analysis, like, for example, face identification or object classification.

Those processed data are then transferred to the third level, i.e., the ``Cloud'', where further analysis, classification, or even federated learning via large-scale models can take place due to the resource's abundance. The three levels of the continuum, along with each one's basic characteristics, are depicted in the diagram in Figure~\ref{fig:Continuum}. For example, data from the highly constrained ``Far Edge'' to the left (e.g., a detected face in a camera frame) travel to the more capable and powerful "Edge" in the middle (trying a face identification against a local database), and eventually to the ultra-powerful and resource-abundant ``Cloud'' to the right for further comparison, e.g., against a database collection of famous persons' images. Then, the ``Cloud'' sends refined results back to the ``Edge'', which then forwards the results to the ``Far Edge'' accordingly.

The Continuum, based on the characteristics of each level, is designed to address and optimize specific AI tasks and address factors like latency, resource availability, and privacy concerns, along with data volume handling~\cite{EdgeAIsurvey2023, arxiv2025edgeSurvey}.

\begin{figure}[!t]
\centering
\subfloat[Multi-Protocol Architecture Implementation and Data Flow.]{
    \includegraphics[width=0.95\textwidth]{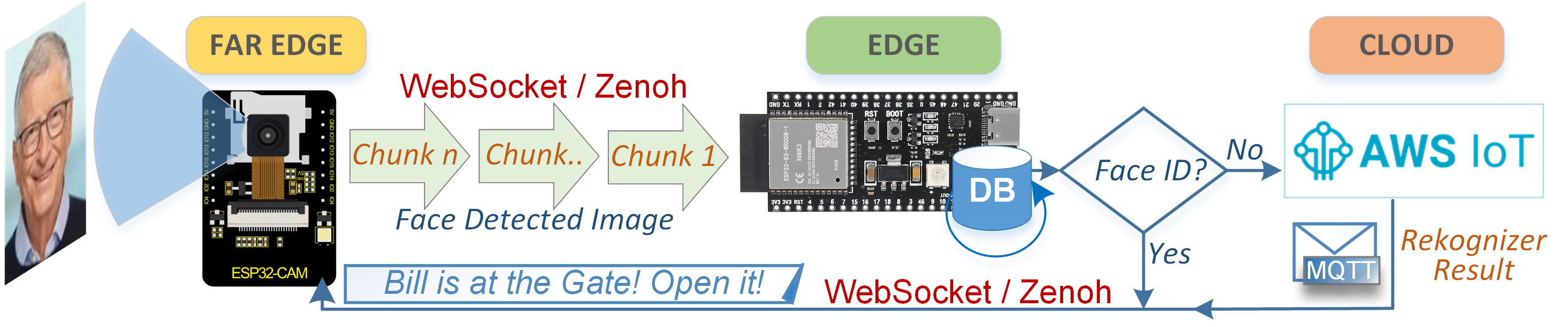}
    \label{fig:hw-data-flow}
}
\vspace{0.5em}
\subfloat[Far Edge Client sends detected face(s) and receives results.]{
    \includegraphics[width=0.48\textwidth]{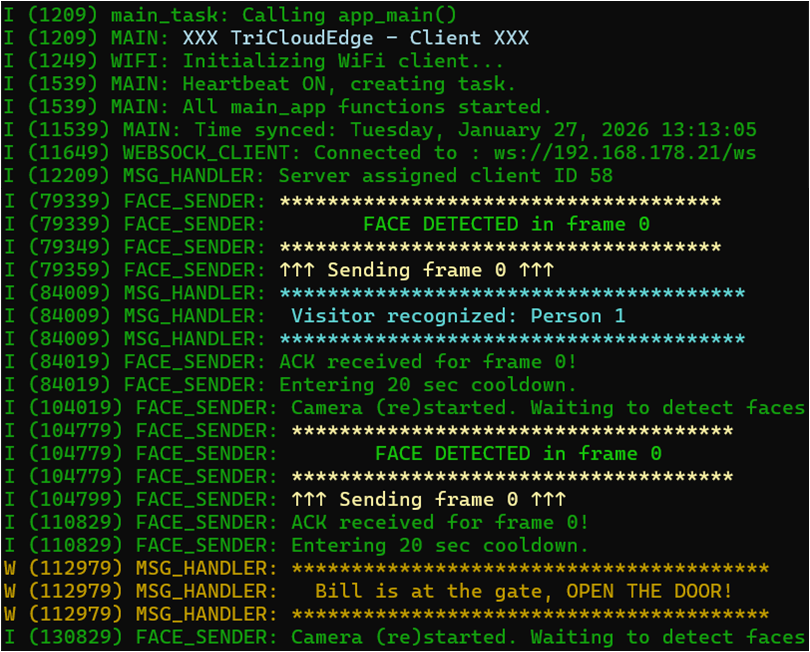}
    \label{fig:client_face_full}
}
\hfill
\subfloat[Edge Server tries face identification, if not, sends to cloud, sends results back to far edge.]{
    \includegraphics[width=0.48\textwidth]{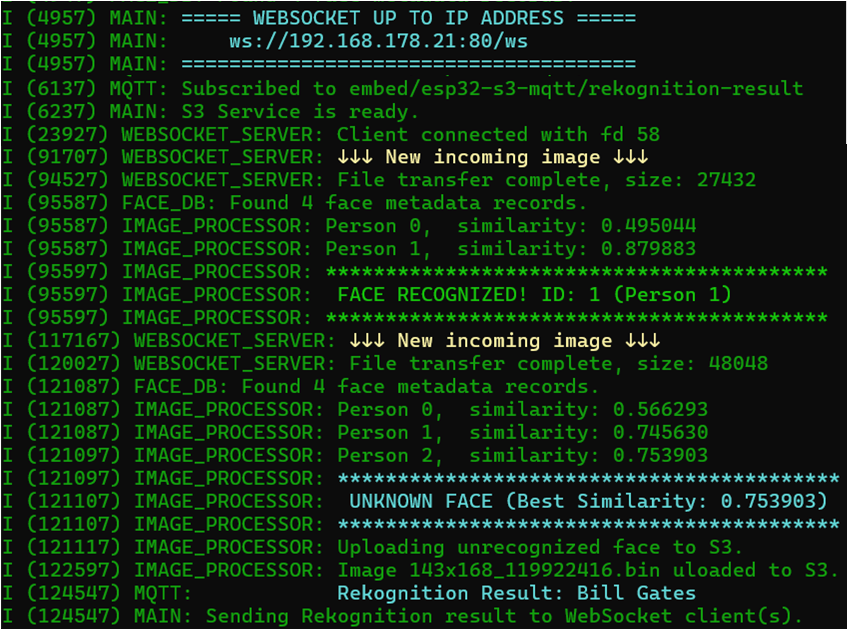}
    \label{fig:server_face_full}
}
\caption{Far Edge - Edge - Cloud Continuum in action.}
\label{fig:dataflow_and_Bill_screenshots}
\end{figure}

\subsection{Far Edge: Constrained but Real-Time}

More specifically, the far edge represents computational nodes with severe limitations in resources and very low power consumption~\cite{powerConsumption025}, as it could potentially be powered by batteries. Those devices target (near) real-time local sensing and event-driven processing with immediate, responsive actions~\cite{tinyML2024}.
The microcontroller ESP32-CAM~\cite{esp32cam,esp32_paper}, a promising candidate for future edge-layer deployments in distributed fog-computing systems~\cite{MARTINEZ2024104026}, also used for embedded video applications due to the integrated camera and the wireless connection capabilities (e.g., Wi-Fi), is a typical example of such a device, costing less than €10.
%
%
Such devices feature memory capacities of 0.5 to 1 MB of RAM and 4 to 16 MB of Flash storage. Any AI tasks executed on this far edge layer are inherently fundamental; for example, immediate object or face detection only~\cite{lowFaceNet2024} (\texttt{FACE DETECTED in frame 0} in Fig.~\ref{fig:client_face_full}). 

At that level, sensors provide raw, primitive data or small-scale images and video frames of low resolution~\cite{tinyML2024}. Preserving the locality of these raw data ensures that personal information never leaves the device unless it is really necessary, preserving privacy and reducing the exposure risk~\cite{tchaye2023privacy}.

\subsection{Edge: Local Intelligence and Enhanced Capabilities}
The middle part of Figure~\ref{fig:Continuum} depicts the Edge Layer, where more capable devices provide intelligence and more processing power~\cite{EdgeAIsurvey2023, meloni2024edgeai}. 
Such examples include the ESP32-S3 board~\cite{esp32s3}, which offers built-in AI acceleration, along with more memory, usually from 0.5 to 8 MB of PSRAM and up to 32 MB of Flash, with a cost of a bit more than €10.
This layer can perform local inference, serious aggregation, and processing while preserving privacy due to the locality of the events. It can also exhibit some tolerance to network connectivity~\cite{meloni2024edgeai}.

Such an edge-processing addresses key hardware constraints as identified in European Union roadmaps~\cite{EPoSSINSIDE_EdgeAI_2025}, as the so-called ``memory wall'' limitation on processing power on those tiers. The same study mentions that future solutions will demand innovation in specialized accelerators (e.g., RISC-V extensions, in-memory computing, and spiking neural networks (SNNs)) to maintain high performance and energy efficiency at the deep edge. Performing inference at the edge also protects sensitive data (e.g., biometrics) by ensuring they remain local. This locality reduces cloud-connectivity dependencies, increases operational resilience, and prevents personal data from propagating unnecessarily to higher layers of the continuum~\cite{tchaye2023privacy}.

Hence, advanced AI tasks can be executed on that level, such as object classification or face identification, compared to the locally stored database, as, for example, indicated by the output \texttt{FACE RECOGNIZED! ID: 1 (Person 1)} in Fig.~\ref{fig:server_face_full}, after it compared with the whole local database (\texttt{Found 4 face metadata records}).
The AI face identification library shows the output of the multiple similarity scores of each database entry (e.g., in Fig.~\ref{fig:server_face_full}, \texttt{Person 0, similarity: 0.495044}). At this level, the data are filtered, along with processed images and video feed~\cite{EdgeAIsurvey2023}; if the face is not locally recognized, it can be sent to the AWS Cloud for further analysis (Fig.~\ref{fig:server_face_full}, \texttt{Uploading unrecognized face to S3}), and eventually the result is sent back to the Far Edge client (e.g., in Fig.~\ref{fig:server_face_full}, \texttt{Sending Rekognition result to WebSocket client(s).}). The result is then received by the Far Edge client, either after sending directly to the Edge the detected person's information (Fig.~\ref{fig:client_websock_aws}, \texttt{Visitor recognized: Person 1}), or, after the result comes back from the AWS cloud, it triggers some actuator(s) (Fig.~\ref{fig:client_face_full}, \texttt{Bill is at the gate, OPEN THE DOOR!}).

\subsection{Cloud: Centralized Analytics on a Global Scale}
The column on the right part of Figure~\ref{fig:Continuum} represents the Cloud part as the more powerful and final part of the Continuum. In this layer, all resources are practically infinite (e.g., processing, storage), provided by, for example, hyperscale datacenters~\cite{hyperscale2023energy}, along with the capabilities to do data aggregation and large-scale AI model testing and training, among other things~\cite{arxiv2025edgeSurvey, 2018nextGenCloud}.

Platforms like AWS IoT Core~\cite{aws} can be utilized here for data aggregation and storage, multiple different big-scale AI models testing and comparison, and possible Internet crawling and novel research, with an initial cost of only a few euros per month.
On that layer, global identity management can take place (e.g., in Fig.~\ref{fig:server_face_full}, AWS sends the face identification result: \texttt{Rekognition Result: Bill Gates}), along with the orchestration of federated learning initiatives and extensive big data analytics~\cite{EdgeAIsurvey2023, li2020federated}.
The storage capabilities of the layer can be measured from terabytes to petabytes. 

\subsection{Computational Power vs Low Latency}

The basic trade-off addressed by the above architecture is the balance between computational power and low-latency. 
Far-edge offers minimal latency and data privacy by eliminating network transfers, but is heavily constrained by limited processing, storage, and energy resources.
The intermediate tier, the edge computing, offers a balance by bringing advanced computation closer to the data source than the Cloud and hence reduces latency for real-time applications~\cite{caiazza2022edge}. 
Some negative aspects of this level are the heterogeneous resources and potential load-balancing challenges~\cite{li2020load_balance}.
Finally, at the classic Cloud tier, platforms like AWS may offer virtually unlimited, scalable computational power and centralized management, targeting complex, computation-intensive processing. However, all this comes at the cost of high and variable network latency, as data must travel via the Internet, which is rather unsuitable for time-sensitive tasks~\cite{prangon2024ai}.

This tiered structure, i.e., the ``Computing Continuum'' allows the optimization of the placement of workload based on the application's requirements, consistent with IoT-based smart-city systems where machine learning tasks are distributed across constrained devices like the ones used in this essay (ESP32), edge nodes, and cloud services to balance energy, latency, and processing limits~\cite{nikpour2025intelligent}. 
For example, in a medical image processing use case, it can bring a balance between cloud-available high-performance Graphics Processing Units (GPUs) versus the low-latency, privacy-preserving processes implemented at the edge~\cite{akdemir2025tomography}.
Simple, time-critical tasks (e.g., immediate threat detection) can be handled at the far edge, while more complex analytics that can tolerate delay (e.g., model retraining) are offloaded to the cloud. The middle edge tier can manage intermediate tasks that require more computation than the far edge device can handle, but cannot tolerate the latency of the cloud.

\subsection{Data and Model Flow}

The continuum operates over bidirectional flows of data and model updates, utilizing the specific characteristics of each layer, targeting a synergy-optimal result as follows (Fig.~\ref{fig:dataflow_and_Bill_screenshots}):
\begin{itemize}
\item Far Edge to Edge: Far Edge devices transmit data, local events, and pre-processed feeds from local sensors to the Edge for refined and advanced processing and aggregation, reducing bandwidth requirements along with lowering the cost of uploading to the (usually data-volume charging) Cloud.
\item Edge to Cloud: The Edge transmits the already aggregated results, comprehensive anomaly reports, and model updates (e.g., Federated Learning) to the Cloud. Such data support large-scale global model training, long-term archival, and advanced big data analytics~\cite{hyperscale2023energy, book2020edge}.
\item Cloud to Edge: The Cloud deploys new trained model versions, global configuration updates, and updated policies to the edge devices, improving local intelligence and operational parameters~\cite{book2020edge, arxiv2025edgeSurvey}, over federated‑learning frameworks~\cite{DJAMAA2025104357}.
\item Edge to Far Edge: The edge forwards lightweight model updates and exact command-and-control signals to the far-edge devices, enabling local adaptations, triggering specific actions, and ensuring the efficiency and process capabilities of the most constrained far-edge devices~\cite{hyperscale2023energy}.
\end{itemize}

This interconnected and interactive feedback architecture ensures the maximum utilization of resources and the efficient and smooth operation of the system by balancing real-time requirements, overcoming system constraints and limitations, addressing privacy concerns, and dynamically adapting with continuous improvements of AI models results~\cite{LatentAI2025EdgeContinuum, arxiv2025edgeSurvey}. 

Furthermore, given the constraints in complexity, latency, and resource utilization of the ESP32 platforms, it is worth studying the possible improvement imposed when shifting from a heterogeneous, multi-protocol stack to a unified, zero-overhead solution, as for example, the Zenoh protocol~\cite{zenoh}.
So, to support a comparative analysis (Section~\ref{sec:comp_analysis}), the continuum was implemented using two distinct communication architectures, i.e., a ``Multi-Protocol Architecture'' and the ``Zenoh-Unified Architecture''. The model and data flow remained identical, executed only over the different transport and communication layers and protocols. 

\subsection{Communication Protocols}

Below, we describe in high-level terms the basic protocols used in this work.
\begin{itemize}
\item WebSocket is a full-duplex communication protocol enabling persistent, bidirectional connections as a server-client model operating over a single TCP socket. As such, the WebSocket protocol eliminates the usual HTTP polling overhead by keeping the connection(s) open and therefore enables real-time data exchange with rather minimal latency. The protocol is used in, for example, web applications where live updates are required, as in gaming, chat, or any kind of dashboard~\cite{websocket}.
\item {MQTT (Message Queuing Telemetry Transport) is a rather new, lightweight protocol based on the publish/subscribe messaging model that targets low-bandwidth, high-latency, and/or unreliable networks. It operates over TCP/IP, optimized for constrained devices and machine-to-machine communication. Among others, it supports persistent sessions and minimal overhead, making it ideal for IoT deployments across domains like industrial, medical, etc.~\cite{mqtt}.}
\item {Zenoh is a newer protocol designed to unify data communication, storage, and computation for cloud-to-device systems. Much like MQTT, it supports pub/sub (publication/subscription) and query patterns. It can potentially operate efficiently over heterogeneous networks without topological constraints, operating over both TCP and UDP with either peer-to-peer or routed communication. Zenoh targets high performance and low-latency, and is optimized for edge computing, IoT, and real-time applications~\cite{zenoh}.}
\end{itemize}

\begin{figure}[t]
  \centering
  \subfloat[Total runtime and throughput with pipeline parallelism enabled.]
  {\includegraphics[width=0.48\linewidth]{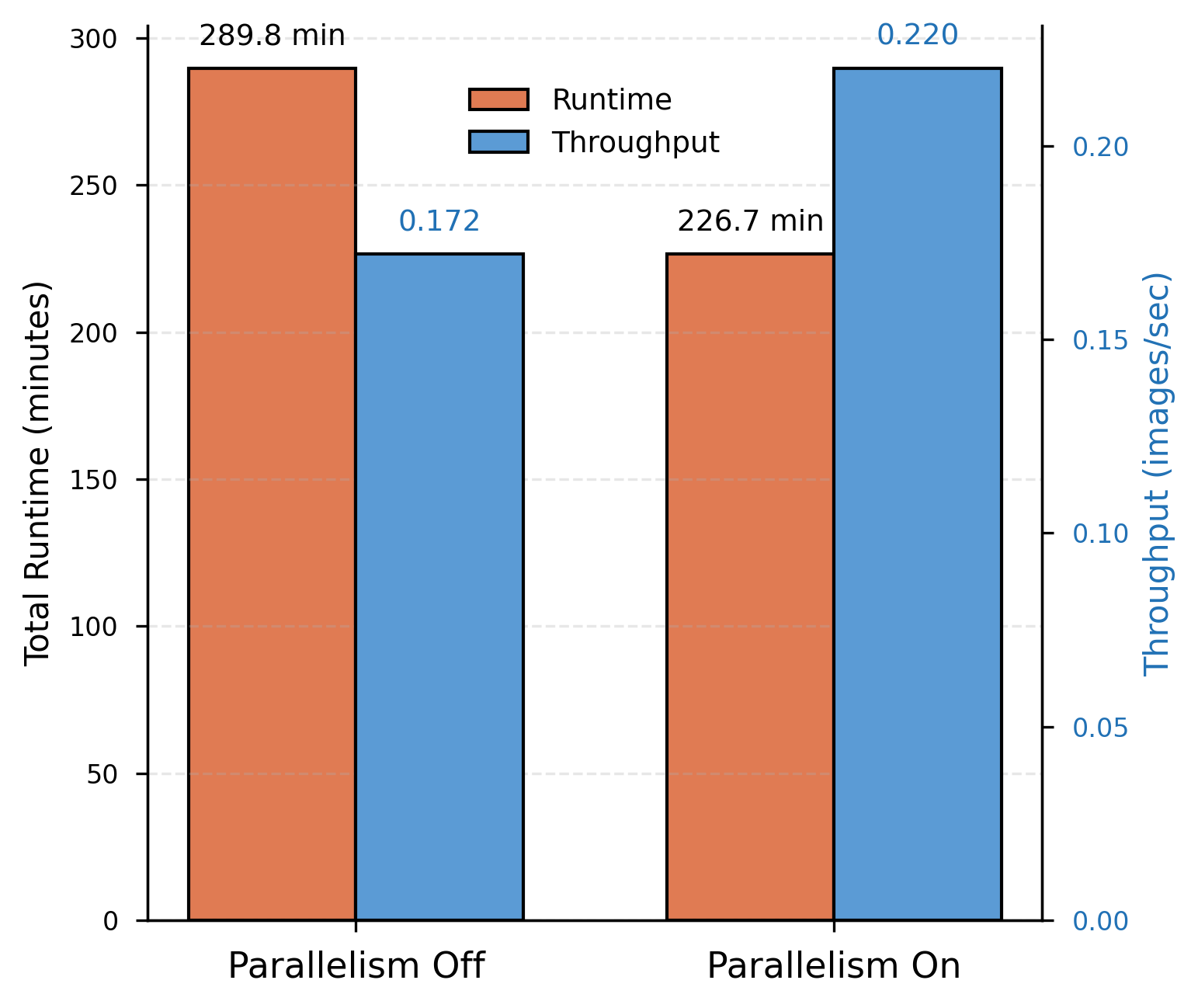}
  \label{fig:runtime_throughput}}
  \hfill
  \subfloat[CPU core utilization (Core 0, Core 1, total) with parallelism enabled.]
  {\includegraphics[width=0.48\linewidth]{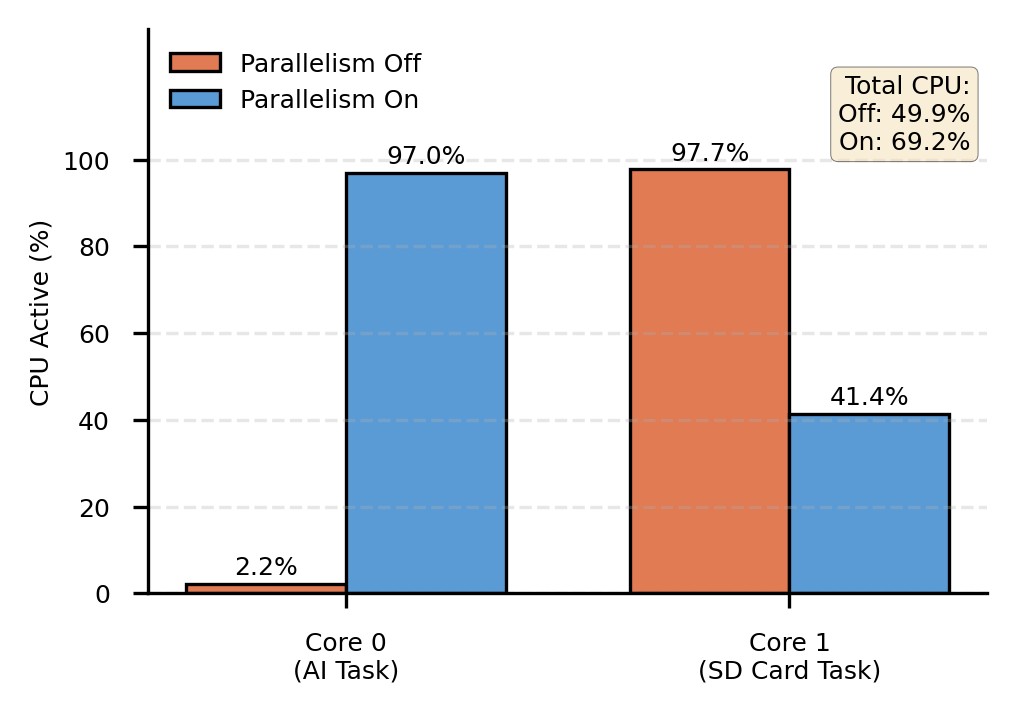}
  \label{fig:cpu_util}}
  \vspace{0.5em}
  \subfloat[Throughput per protocol with parallelism enabled.]
  {\includegraphics[width=0.48\linewidth]{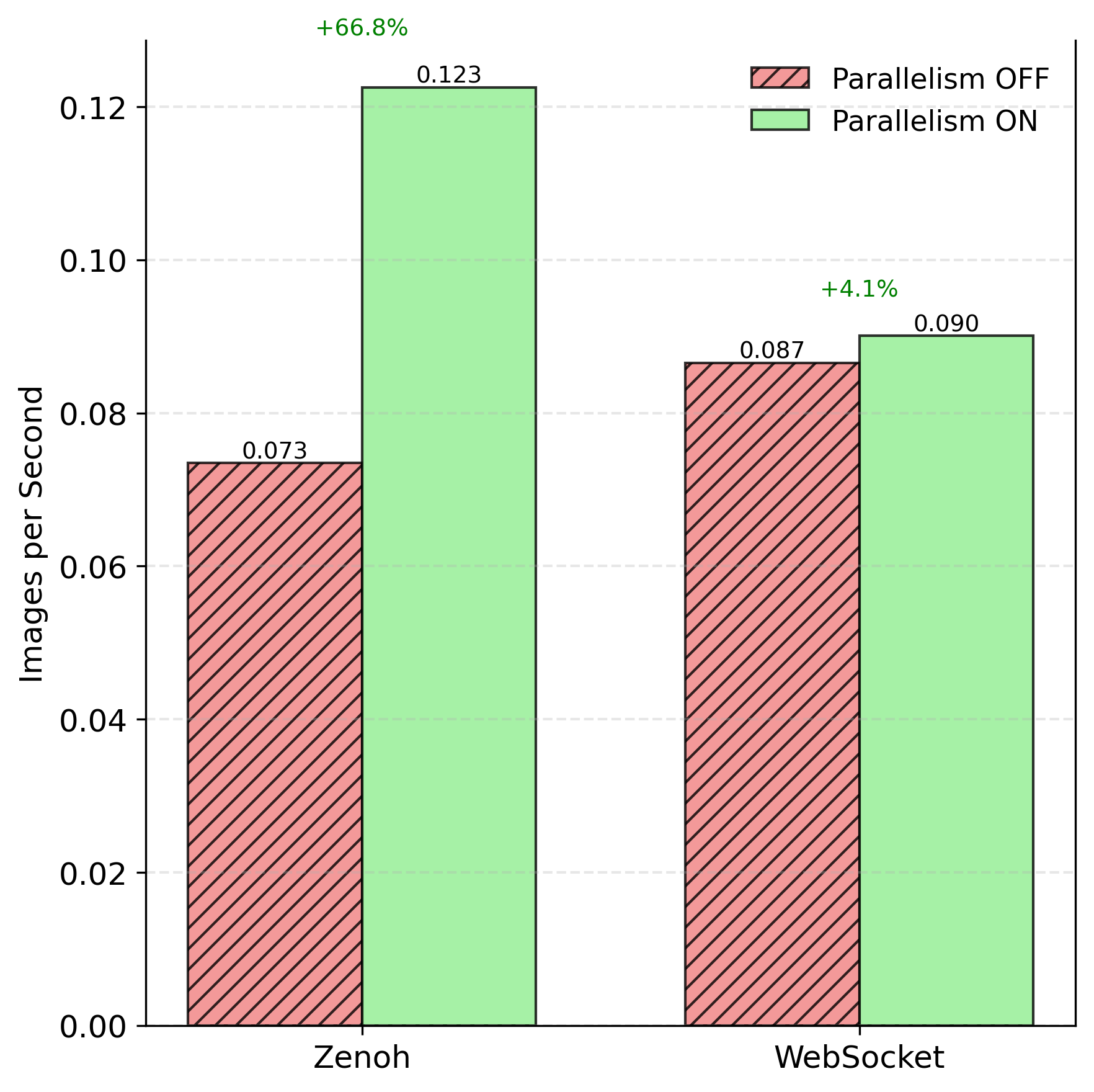}
  \label{fig:process_parl}}
  \hfill
  \subfloat[Runtime and Throughput improvement from enabling parallelism.]
  {\includegraphics[width=0.48\linewidth]{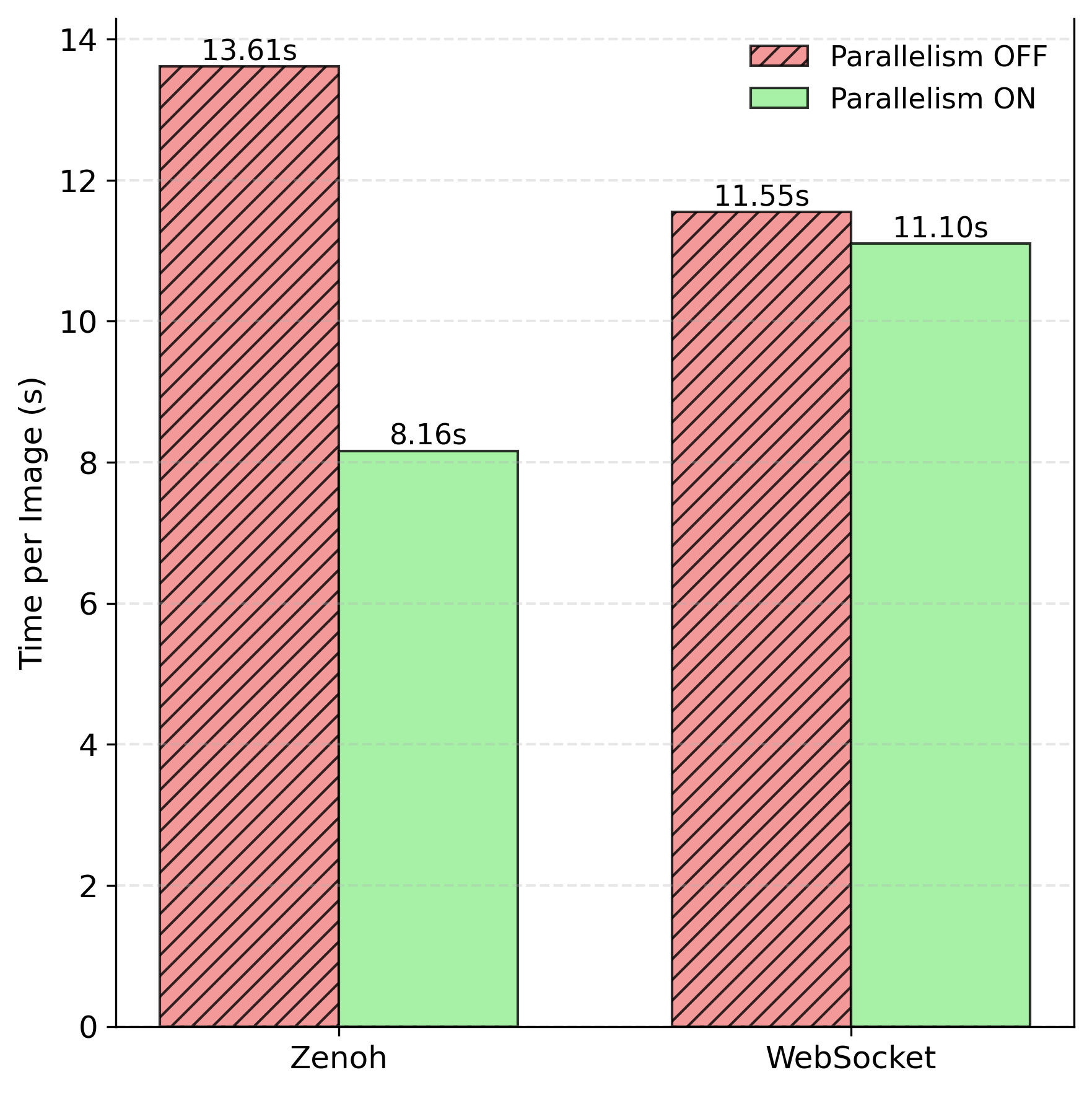}
  \label{fig:throughput}}
  \caption{ESP32-CAM pipeline parallelism and per-protocol performance over a 3,000 images dataset.}
  \label{fig:runtime_vs_cpu_util_parl_on_off}
\end{figure}

\section{Comparative Analysis}
\label{sec:comp_analysis}

We hereby present the experimental evaluation of the TriCloudEdge Continuum, focusing on (i) the parallelism impact on constrained devices like EPS32-CAM, (ii) latency and throughput between WebSocket and Zenoh, and (iii) some insights on the architectural complexity imposed by different approaches, such as multi-protocol stacks versus a more unified approach. All experiments use the same hardware, software, and AI stacks described in Section~\ref{sec:system_architecture}.

\subsection{Methodology}

\subsubsection{Experimental Setup}

The evaluation deploys the three-layer continuum described in Section~\ref{sec:system_architecture}, over the following hardware and software stack:

\begin{itemize}
    \item Far Edge: ESP32-CAM, performing on-device face detection over locally stored or streamed image frames
    \item Edge: ESP32-S3 with PSRAM, executing face identification against a local face database and communicating with the Cloud
    \item Cloud: AWS components, including IoT Core, S3, Lambda functions, and ``Rekognition'' for large-scale analytics and global identity management
\end{itemize}

All embedded components are implemented using the ESP-IDF toolchain with identical compiler settings. The same face-detection and face-identification models were used in all experiments, and the same custom fragmentation protocol was used for both protocols (more implementation details in Section~\ref{sec:implementation_details}). On the cloud side, we use the same AWS configuration and identical Lambda/Rekognition implementations for all evaluated scenarios. 
All the project links provided in Section~\ref{sec:data_and_code} have detailed reproducibility lists and instructions (e.g., hardware versions, ESP-IDF libraries versions, and configuration options like SPIFFS setup), along with all data used in the experimentations.

\subsubsection{Evaluation Metrics}

The following metrics were utilized for evaluation:

\begin{itemize}
    \item End-to-end latency, from far-edge face detection to the final response delivered back to the far edge (Edge-only or Edge+Cloud)
    \item Protocol latency and throughput, including per-stage round-trip times (RTTs) and overall application runtime
    \item CPU utilization and task-level runtime on the embedded devices
    \item Memory footprint (RAM/Flash) and binary size on constrained devices
    \item Model performance under constrained execution, in terms of feasibility and processing time
\end{itemize}

\subsubsection{Experimental Procedure}

The typical end-to-end flow is as follows:

\begin{itemize}
    \item The Far Edge (ESP32-CAM) performs local face detection on incoming frames and transmits detected faces to the Edge using the designated protocol (WebSocket or Zenoh)
    \item The Edge (ESP32-S3) performs face identification against a local database. If the face is recognized, the result is either sent immediately back to the Far Edge or to the cloud for further analysis. The cloud response is then propagated to the far edge
    \item Each protocol is evaluated under identical conditions and datasets 
    \item Mean and variance of per-stage timings and overall runtimes were collected
    \item  End-to-end latency measured from the initial detection (Far Edge) til the final response received (from Edge or from the cloud)
\end{itemize}


\paragraph{Configurations evaluated}

The evaluation covers the configurations listed in Table~\ref{tab:configurations-evaluated}. Each figure in this section corresponds to a specific configuration as noted in the table and in the figures' captions. 

\begin{table}[t]
\centering
\caption{Configurations evaluated and their mapping to figures.}
\label{tab:configurations-evaluated}
\scriptsize
\setlength{\tabcolsep}{3pt}
\resizebox{0.48\textwidth}{!}{%
\begin{tabular}{@{}c c c c c p{1.4cm}@{}}
\toprule
\textbf{Fig.} & \textbf{Dataset} & \textbf{AWS} & \textbf{Parl.} & \textbf{Prot.} & \textbf{Metric(s)} \\
\midrule
\ref{fig:runtime_vs_cpu_util_parl_on_off} & 3k & OFF & ON/OFF & Z, W &
Runtime, thruput, CPU, gain \\
\ref{fig:cloud_overhead} & 47 & ON & ON & Z, W &
Edge vs.\ Cloud RTT \\
\ref{fig:multi_dataset_compare} & \textbf{Multi} & ON/OFF & ON & Z, W &
Thruput vs.\ dataset \\
\ref{fig:client_server_screenshots} & 47 & ON/OFF & ON & both &
Terminal (6a, 6b) \\
\bottomrule
\end{tabular}%
}
\vspace{0.5em}
\raggedright
\scriptsize
\textbf{Notes:}\\
\textbf{Fig.}: figure in this paper\\
\textbf{Dataset}: Number of AI images used (links in Section~\ref{sec:data_and_code})\\
\textbf{Multi}: AI images dataset sizes: 47, 100, 150, 200\\
\textbf{AWS}: Edge node forwards images to AWS/Rekognition and receives results \\
\textbf{Parl.}: far-edge parallelism \\
\textbf{Prot.}: WebSocket (W), Zenoh (Z)\\
\end{table}

\subsection{Pipeline Parallelism and Per-Protocol Performance}
\label{sec:paralell_vs_performance}

The effect of enabling pipeline parallelism on the ESP32-CAM is evaluated below; possible results from different angles and trade-offs are presented in Fig.~\ref{fig:runtime_vs_cpu_util_parl_on_off} to show the aggregate gain and per-protocol behavior. The far edge device processes a local database of images stored on the SD card, performing PNG file reading, decoding, and two-stage face detection on each image. Pipeline parallelism distributes I/O and AI processing across the two CPU cores, overlapping SD card access with inference.

The overall impact of parallelism is depicted in Fig.~\ref{fig:runtime_throughput}, where, with parallelism enabled, a substantial performance improvement is gained. More specifically, for the continuous processing of 3,000 image files from the SD card, the total runtime is reduced by ~21.8 percent (from 289.8 to 226.7 minutes), or in a different view, also in Fig.~\ref{fig:runtime_throughput}, the throughput increases by ~27.9 percent (from 0.172 up to 0.220 images per second).

The CPU utilization is depicted in Fig.~\ref{fig:cpu_util}. Without parallelism, one core remains mostly idle while the other is almost saturated (overall utilization as [2.2 + 97.7] / 2 =  49.9). With parallelism enabled, the core where AI is running goes up to 97.0 percent, while the core doing the I/O goes up to 41.4 percent, yielding a better-balanced CPU utilization of 69.2 percent.

Figure~\ref{fig:process_parl} shows the throughput by protocol (Zenoh vs WebSocket) when parallelism is enabled. Given the modern, up-to-date architecture of Zenoh, it benefits heavily from enabling parallelism (a gain of +66.8 percent), while WebSocket marginally improves by +4.1 percent. Furthermore, Fig.~\ref{fig:throughput} depicts the per-protocol improvement by enabling parallelism, expressed as the total runtime reduction, or throughput improvement.

Those results confirm that even in highly constrained devices, enabling pipeline parallelism can significantly improve overall utilization of resources and improve throughput without even modifying the AI models utilized. 
A detailed analysis and discussion of per-image processing time is provided in Section~\ref{sec:stat_analysis_process_time}.

\begin{figure}[!t]
\centering
\includegraphics[width=0.48\textwidth]{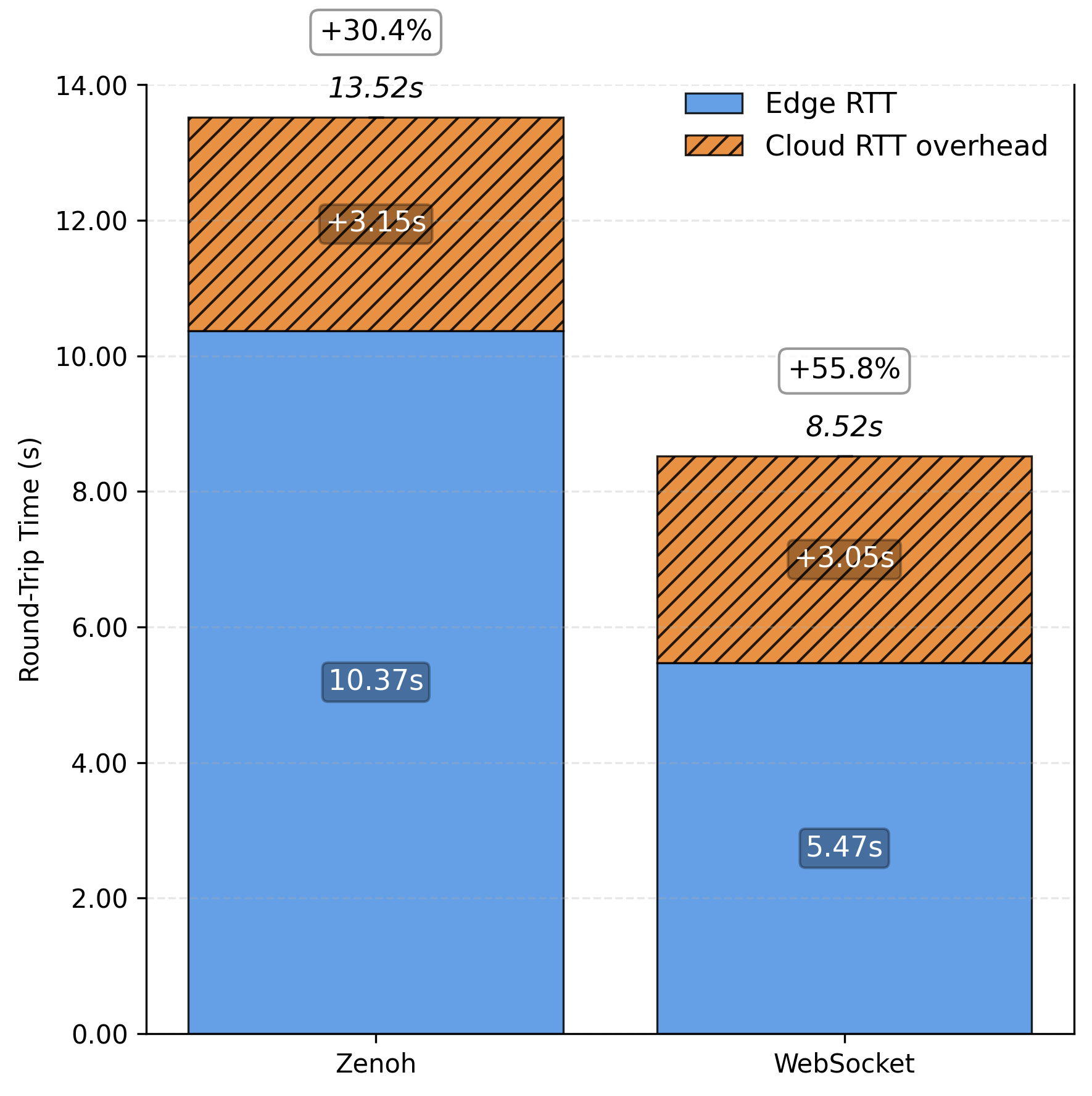}
\caption{Comparison of RTT to Edge vs Cloud (AWS).}
\label{fig:cloud_overhead}
\end{figure}

\subsection{Protocol Performance: Latency and Throughput}
 \label{sec:protocol_performance_latency_vs_throughput}

Here, we compare the performance of the two communication architectures, defined as the Multi-Protocol Architecture based on WebSocket, HTTP and MQTT as main components, and the Zenoh-Unified Architecture, where Zenoh is solely responsible for the communication between the far edge and edge tiers. All AI workloads and the fragmentation (chunking) protocol remain identical.

\begin{figure}[!t]
\centering
\includegraphics[width=0.48\textwidth]{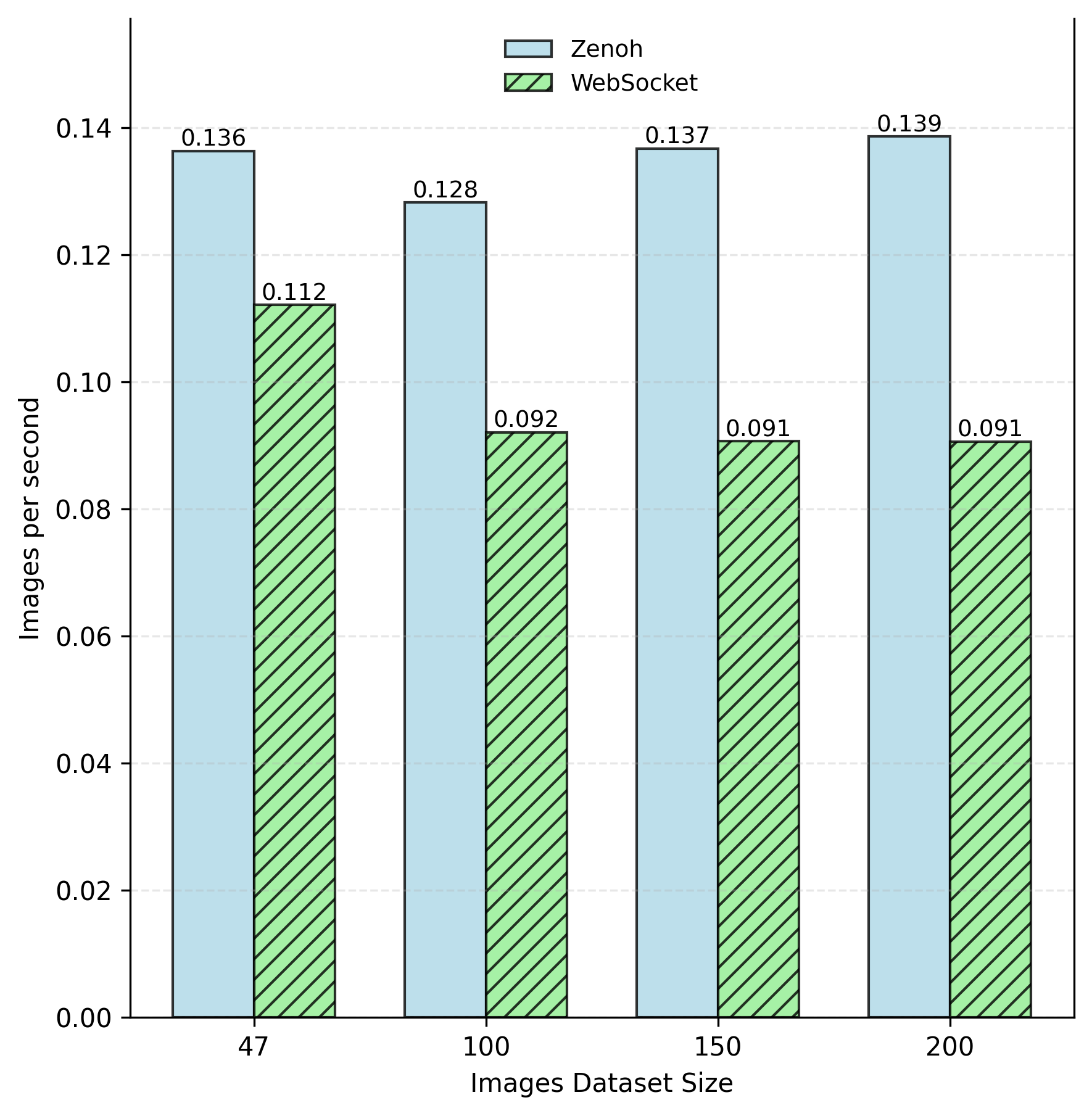}
\caption{Comparison of protocol throughput over multiple datasets.}
\label{fig:multi_dataset_compare}
\end{figure}
\begin{figure}[!t]
\centering
  \subfloat[ESP32CAM terminal output. Detected faces are sent to Edge for further analysis.]
  {\includegraphics[width=0.48\textwidth]{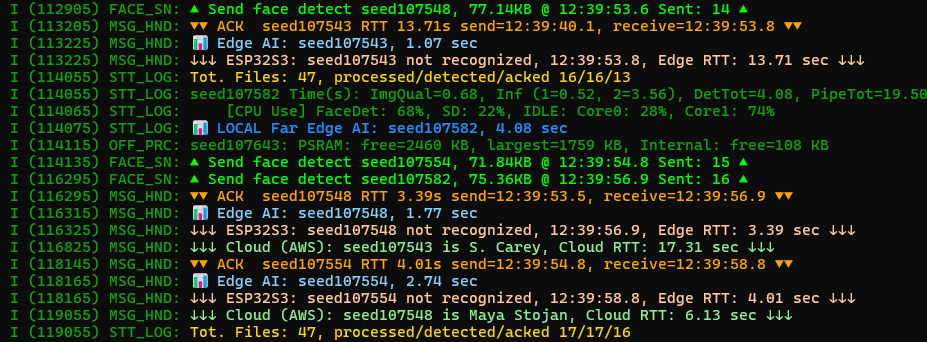}
  \label{fig:client_websock_aws}}

  \subfloat[ESP32-S3 terminal output. Images with faces not identified locally are sent to the cloud.]
  {\includegraphics[width=0.48\textwidth]{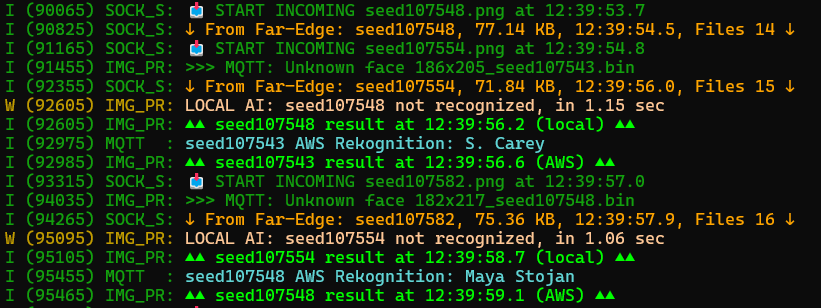}
  \label{fig:server_websock_aws}}
  \caption{Far Edge - Edge devices in action. Images are sent to AWS, and results are sent back to Far Edge.}
  \label{fig:client_server_screenshots}
\end{figure}

\begin{enumerate}
    \item End-to-End Latency:

    Figure~\ref{fig:cloud_overhead} shows the end-to-end latency for both architectures, from edge-only to the edge+cloud round trip. Adding the cloud layer increases the latency by an average of 25 percent since the image has to travel to AWS and execute the Lambda/Rekognition there. For both architectures, the cloud latency is almost identical, as expected, since that part is identical for both.


    Far edge face detection (which includes image quality assessment and two-stage inference by the ESP embedded AI engine) takes about 4.45 seconds per image, while the edge-side face identification takes about 1.09 seconds per image (such indicative times are depicted in Fig.~\ref{fig:client_websock_aws}, for example, \texttt{LOCAL Far Edge AI: seed107582, 4.08 sec}, and in Fig.~\ref{fig:server_websock_aws}, \texttt{LOCAL AI: seed107554 not recognized in 1.06 sec}. The only difference is the protocol-dependent stage, i.e., the image uploading from far edge to edge, where, with parallelism enabled, Zenoh exhibits a mean of 8.16 seconds, while the WebSocket exhibits 11.10 seconds (Fig.~\ref{fig:throughput}).


    \item Throughput and Multi-Hop Behavior:

    While per-file end-to-end latency seems to be better for WebSocket (Fig.~\ref{fig:cloud_overhead}), the effective throughput over multiple datasets measured as the images per second in Fig.~\ref{fig:process_parl} or the per-file RTT in Fig.~\ref{fig:throughput} (e.g., 13.61 versus 11.55 seconds for enabled parallelism) are both in favor of Zenoh, indicating the better usage of parallelism from Zenoh's asynchronous nature against the blocking monolithic characteristics of WebSocket. 
    
 \item Architectural Complexity:
 
    The Zenoh-Unified Architecture uses a single API and Zenoh's data-centric model to handle pub/sub and storage, reducing the number of abstractions and the need for extra protocols. Considering the improvement in throughput results, as depicted in Figures~\ref{fig:cloud_overhead} and~\ref{fig:multi_dataset_compare}, if the Zenoh-Unified Architecture is applied end-to-end, including implementation into the cloud, it can offer superior performance and simpler integration and maintenance. The higher per-file transfer time is not usually faced in real-life applications, where, for example, the camera just captures images periodically and sends one at a time for further analysis.

\end{enumerate}

\section{Implementation Details \& Challenges}
\label{sec:implementation_details}

During the experimental evaluation of the project, several practical challenges surfaced while trying to deploy the three-tier cloud continuum with heterogeneous protocols and applications over such constrained devices as the ESP32-CAM and the ESP-S3. Hence, some main issues faced and engineering decisions taken at each layer, as well as the justification of statistics and measurements, are discussed below.

\subsection{Far-Edge Implementation Challenges}
\label{sec:far_edge_challenges}

The ESP32-CAM, as a constrained device, has limited RAM and Flash, no hardware accelerators, and no memory defragmentation mechanism. Hence, repeated image captures and consequent allocations and de-allocations led to memory fragmentation, leading to repeated buffer overflows and stack corruption. To address this, a custom ``chunking'' protocol was created, where each image was split into small segments (1 KB proved to be an effective choice), and along with accompanying metadata, encapsulated into JSON messages. Obviously, this leads to some protocol overhead, yet the protocol was used for both protocols. Furthermore, Zenoh was configured for peer-to-peer over TCP/IP for a fair comparison with WebSocket.

However, Zenoh showed further limitations and issues. In particular, it required LwIP configuration adjustments and coordination with the protocol's engineers to bypass error-recovery issues and prevent unrecoverable failures (not entirely successfully). Furthermore, because of the ESP32-CAM's lack of a memory defragmentation mechanism, running Zehon in full asynchronous mode proved problematic. On the other hand, WebSocket operates in a more serial, blocking way; hence, it proved more robust over the far edge constrained device.

\subsection{Edge Implementation Challenges}

Over the Edge tier, the ESP32-S3 had to host both the local face-identification AI embedded application and the networking stack, along with a local database of face images (for local face identification, without connecting to the cloud). To satisfy the memory demands arising from the above, PSRAM prioritization and SPIFFS expansion were enabled via the ESP-IDF ``menuconfig'' interface. Such a configuration allows the device to store and process a sufficient number of locally stored files (images) if local identification is enabled and leaves enough memory space for necessary protocols and applications.

The verbosity of the ESP-IDF's logging mechanism had to be carefully adjusted, since excessive logging easily led to overwhelming the available memory and introducing delays or even frequent crashes.

\subsection{Cloud Implementation Challenges}

The cloud aspect includes the AWS services for handling unrecognized faces and demonstrates the scalability of the Continuum. AWS Lambda functions were created to receive image data in RGB565 frames. Then, utilizing libraries like NumPy and Pillow, the images were decoded and reconstructed, planned in such ways to avoid runtime bottlenecks.  

In addition, IAM roles and permissions had to be fine-grained in order to securely connect to services like S3, Rekognition, and IoT Core. The pre-signed URLs provided by S3 allow the devices to upload the data, yet they need synchronization between the Lambda function and the device calls. These points are one of the major arguments in favor of proposing the simplification of the overall protocol stack (maybe introducing Zenoh in AWS), as discussed in Section~\ref{sec:future_steps_and_discussion}.

It is worth mentioning that the ``Celebrity Rekognition'' of AWS seems to introduce a high number of false positives, up to 62.7 percent (e.g., for 158 AI-created images sent, 99 of those were identified as some famous person, e.g., Fig.~\ref{fig:server_websock_aws}, \texttt{seed107543 AWS Rekognition: S. Carey}, \texttt{seed107548 AWS Rekognition: Maya Stoyan}), presumably from fluctuating confidence scores, and hence reveals the need for further post-processing filters and threshold adaptation~\cite{probabilistic2004}.

\subsection{Statistical Analysis and Justification of Processing Time}
\label{sec:stat_analysis_process_time}

To extract and validate measured runtimes and statistics, the FreeRTOS runtime statistics were utilized. 

The analysis shows that, on the ESP32-CAM, the face-detection inference task is the superior consumer of the CPU time (approximately 97 percent of the core where it runs), while the data preprocessing task (including file I/O and image decoding from the SD card) takes a smaller but still significant share (on the order of 9 percent). All other tasks, including networking (e.g., Wi-Fi) and event handling, consume less than 2 percent of the total CPU time and are therefore not regarded as dominant contributors to the end-to-end latency.

The average per-image processing time at the Far Edge was measured as approximately 4.8 seconds (e.g., Fig.~\ref{fig:client_websock_aws}, \texttt{LOCAL Far Edge AI: seed107582, 4.08 sec}). This number includes the combined cost of PNG file reading, decompression, and decoding (e.g., with the LodePNG library), image resizing, color conversion from RGB888 to RGB565, and the two-stage AI inference pipeline (e.g., utilizing MSR01 and MNP01 models, as provided by the ESP-IDF embedded libraries). All those operations are software-executed on the EPS32-CAM's dual-core Xtensa LX6 processor, at 160 MHz, with 4 MB external PSRAM, and no dedicated AI accelerators (hence the face detection only). As such, the observed processing time matches the expected performance of such devices as indicated by the vendor~\cite{espressif2023espdl}.

On the edge side, for ESP32-S3, the cost of face identification and database comparison is approximately 1.1 seconds (e.g., in Fig.~\ref{fig:server_websock_aws}, \texttt{LOCAL AI: seed107554 not recognized, in 1.06 sec}), significantly lower than the cost of initial face detection at the far edge. Consequently, offloading the detection to the Far Edge tier and performing identification on the Edge tier proves the justification of such a trade-off, affecting the total latency, especially combined with utilizing the parallelism of the devices and network protocol optimization described in previous sections.

\section{Data and Code Availability}
\label{sec:data_and_code}

The full implementation code used in this study is publicly available on GitHub at the following repositories (NOTE: Those are the early versions of the project code. They will be updated upon acceptance):
\begin{itemize}
    \item \url{https://github.com/georgevio/TriCloudEdge}
    \item \url{https://github.com/georgevio/TriCloudEdge-RTT}
    \item \url{https://github.com/georgevio/ESP32-Zenoh}
\end{itemize}

Additionally, all AI-created images, collected from multiple sources, are accessible at:
\begin{itemize}
    \item \url{https://doi.org/10.5281/zenodo.18429687}
\end{itemize}

\section{Future Steps and Discussion}
\label{sec:future_steps_and_discussion}



In this project, privacy is addressed by processing and data localization. A possible expansion would be a framework to handle and manage trust and identity across the heterogeneous entities of the continuum (i.e., Far Edge–Edge–Cloud). Such a layer, as a decentralized data governance, would exploit technologies like Distributed Ledger Technology (DLT) or Decentralized Identity (DID) models to manage access control and provenance for sensitive data (e.g., face recognition logs) that are being inter-exchanged across the continuum~\cite{blockchain25}. It would be able to aggregate, delete, or process data, providing compliance and auditability with, for example, the upcoming CRA (Cyber Resilience Act). A further direction could be formalizing locality‑preserving data flows as a core security and compliance mechanism, ensuring that personal data are processed and retained at the lowest possible tier, again in accordance with GDPR and CRA. 

Some future research questions may include how to dynamically place and optimize the models used across the continuum~\cite{2023dynamic_edge}, or how to adapt and optimize the fragmentation protocol used in both cases, either dynamically or by targeting specific protocol or sensor characteristics. Such challenges imply the need for further consideration of the balance between resource constraints, responsiveness, integrity, quality of service, and cloud orchestration over such edge-AI deployments~\cite{2022continuum}.  


The current evaluation does not measure energy consumption; given that far-edge devices are battery-operated or energy-constrained anyway, future research should involve the aspect of energy-efficiency modeling in the offloading decision process~\cite{lolipopIoT}. While the work at hand explores the dynamics of unifying protocols, the energy saving by avoiding cloud offloading or optimally delegating tasks remains to be quantified, or even the developing dynamic energy models predicting energy consumption~\cite{song2025dynamic}. 

Moreover, we need research on learning-based offloading algorithms, targeting the overall system energy consumption, given specific latency constraints~\cite{EnergyEfficientOffLoad}. Such an adaptive approach is in full alignment with energy preservation requirements by EU~\cite{EPoSSINSIDE_EdgeAI_2025, easy2025}. 
Further investigation of Hierarchical Federated Learning models across the continuum can delegate the model training functions to the edge while maintaining localized inference to the highly constrained far edge, ensuring privacy by keeping the data local and minimizing transfer overhead since it requires transferring only model updates between tiers~\cite{multiFederate2025}.

The current architecture utilizes standard security protocols, such as TLS over WebSocket and MQTT, and the same applies to Zenoh. Future works should solely focus on implementing Post-Quantum Cryptography (PQC), something rather challenging, especially for the far-edge highly constrained devices, which, being in the open, are the most vulnerable. As such, they demand novel, resource-optimized solutions (e.g.,~\cite{liteQSign}), both on hardware and software levels, to ensure that security and latency are both satisfied, especially for sensitive IoT applications (e.g., industrial or security installations). 

To ensure security and integrity across the continuum remain intact, particularly regarding functional safety, we need to prevent physical intrusion and unauthorized access while guaranteeing the integrity of AI models and data during the transfer between edge devices and retraining servers~\cite{EPoSSINSIDE_EdgeAI_2025}. 
Another valuable addition would be the implementation of a data governance layer, utilizing technologies like Distributed Ledger Technology (DLT) and/or Decentralized Identity (DID) to address the transfer of sensitive data (such as face recognition) across the continuum. Such a model will dictate the specific tier that processes the data (aggregate, delete, etc.) to comply with given rules and regulations.

Beyond the technical comparison in this paper, the currently highly fragmented European ecosystem requires strategic focusing on modularity, interoperability, and establishing common standards to avoid vendor interlocking. Such initiatives are essential to bolster the EU's position towards targeting digital sovereignty in the landscapes of semiconductors and AI technologies~\cite{EPoSSINSIDE_EdgeAI_2025}.

\subsection{Lessons learned}
\label{sec:lessons_learned}

We briefly present here lessons learned from the design, implementation, and evaluation of the TriCloudEdge continuum. The following points summarize practical takeaways for developers and architects wanting to deploy a multi-tier edge-cloud system with constrained devices and adapted communication and protocol stacks.

\begin{itemize}
    \item 
When total runtime and throughput matter more than per-file latency, a Zenoh-like architecture is beneficial under the conditions of this evaluation (see, for example, Figures~\ref{fig:runtime_vs_cpu_util_parl_on_off} and~\ref{fig:multi_dataset_compare})
    \item 
In cases where individual file transfer time is critical, and the workload is low, monolithic protocols like WebSocket may be simpler and sufficient
    \item 
Pipeline parallelism on the far-edge device makes a big difference, even when processing files locally (e.g., Fig.~\ref{fig:runtime_vs_cpu_util_parl_on_off})
    \item 
Image fragmentation (e.g., in 1 KB chunks) seems unavoidable on constrained far-edge devices to prevent buffer overflows (Section~\ref{sec:far_edge_challenges})
    \item 
A versatile multi-level protocol like Zenoh can reduce integration and maintenance effort compared to more complex multi-protocol architectures (WebSocket, HTTP, MQTT, etc.), as described in Section~\ref{sec:protocol_performance_latency_vs_throughput}.
\end{itemize}

\section{Content Generated by AI}

AI tools such as Copilot, Gemini, Cursor, and Grammarly were used in the preparation of this manuscript for editing, grammar checking, and knowledge retrieval. No text was copied without an extensive and thorough review by the authors, and all substantive ideas, analyses, and conclusions are the product and sole responsibility of the authors. Additionally, the listed AI tools were utilized for code development, refactoring, and comment creation, along with early paper reviews.

\section{Acknowledgments}

\noindent The surviving author wishes to formally and profoundly acknowledge the foundational and invaluable contributions of Professor Lefteris Mamatas to this work and to the entirety of his Ph.D. research in the past. 

On a personal note, Lefteris was not only my Ph.D. supervisor in the past, but he was a friend and a mentor above all. His mild character and eternal smile made me a better person. I was lucky to have met him, and he will be greatly missed. \foreignlanguage{greek}{Λευτέρη, θα σε θυμάμαι πάντα!}

\bibliographystyle{IEEEtran}
\bibliography{B}

\end{document}